\documentclass[12pt
]{article}
\usepackage{final}

\newcommand{\two}{\{0, 1\}}
\usepackage{float}
\begin{document}

\maketitle

\section*{Abstract}
Witness encryption using multilinear maps was first proposed in 2013 \cite{witnenc_garg13}, and has continued to evolve since. In this paper, we build on an open-source multilinear map implementation by Carmer and Malozemoff \cite{codeclt13} of the graded encoding scheme proposed CLT13 \cite{clt13}, with asymmetric modifications \cite{clt13asym}. Using this map, we created the world's first ciphertext encoded with a candidate witness encryption scheme. Finally, using a reduction from Sudoku to \textsc{Exact Cover}, we encrypted the private key to a Bitcoin wallet with 22,700 Satoshi using a Sudoku.

\section{Introduction to Witness Encryption}
Witness encryption for an NP language $L$ (with witness relation $R$) involves allowing
decryption of a message, that has been encoded with respect to some problem instance $x$, if $x \in L$ and one has a \emph{witness} $w$ for which $R(x, w) = 1$ (\cite{witnenc_garg13}). 
Along with applications in creating cryptographic primitives
\cite{witnenc_garg13}, this allows us to encode monetary rewards
or messages for those who are able to solve a puzzle or problem. 
To cite a famous example, consider the Riemann Hypothesis. Perhaps
the Clay Mathematics Institute would like us to encrypt a key to a
cryptocurrency wallet worth \$1 million, retrievable only by those
who possess a correct proof to the Riemann Hypothesis. 
Using witness encryption, the key to the wallet could be encrypted
using an encoding of the Riemann Hypothesis in some formal language,
with decryption possible if and only if a proof of the Riemann
Hypothesis is presented.

Note that the encrypting party themselves may not know whether or
not a solution exists \cite{witnenc_garg13}. This fact makes 
witness encryption promising for problems of which the solution 
to is unknown, but can be verified quickly. 

\begin{definition}[Witness Encryption by Garg et al. \cite{gargslides}]
We define witness encryption for an NP language $L$ (specifically, solutions to the \textsc{Exact Cover} problem). Since the problem is in NP, we have a circuit $C$ that, given a witness, can determine whether some $x$ is in $L$ or not. From here, our witness encryption scheme comprises of two PPT machines $(Enc,Dec)$. We will encrypt a single bit $m\in\{0,1\}$.
\begin{enumerate}
    \item $\Enc(C,m)$ outputs a cipher text $e\in\{0,1\}^*$.
    \item $\Dec(C,e,w)$, where $w$ is a witness. Our decryption box outputs
    $m$ if $C(w) = 1$ and $\perp$ otherwise.
\end{enumerate}
We want these machines to satisfy two standard conditions.
\begin{itemize}
    \item \textbf{Correctness:} Given that $\exists w$ such that $C(w) = 1$, we have $Dec(C,e,w)= m$.
    \item \textbf{Security:} If $\nexists w $ such that $C(w)=1$, then
    $Enc(C,0)$ is computationally indistinguishable from $Enc(C,1)$
\end{itemize}
\end{definition}
As Garg et. al notes, the security definition of witness encryption only requires indistinguishability of encryptions when the language $L$ is empty (i.e. the problem is unsolvable). When $L\neq \varnothing$, this definition does not guarantee that an adversary with no knowledge of a witness would not be able to decrypt the ciphertext in polynomial time. To that end, there exist stronger definitions of witness encryptions, such as \emph{extractable witness encryption}, which give security guarantees in the solvable case.


\section{Witness Encryption Scheme}

\textbf{Encryption Scheme for \textsc{Exact Cover} from Garg et al. \cite{gargslides}}
Consider the \textsc{Exact Cover} problem, where
given some $S_1,\ldots,S_\ell \in \mathcal{P}([n])$, we want to
find an index set $T\subset [\ell]$ such that $\bigcup_{i\in T} S_i = [n]$ and
$S_i \cap S_j = \varnothing$ for all $i\neq j$.
\begin{itemize}
    \item $\Enc(x,m)$. Takes as input $x=(n,S_1,\ldots,S_\ell)$ and bit $m \in \{0,1\}$. In order to encrypt this bit, perform the following 
    \begin{itemize}
        \item Sample $a_0,\ldots,a_n$ uniformly and independently from $\Z_p^*$.
        \item Let $c_i = g_{|S_i|}^{\prod_{j\in S_i}a_j} $ for all $i = 1,\ldots, \ell$. We are encoding the value
        $\prod_{j\in S_i}a_j$ at \emph{level} $|S_i|$. This notion of levels will be detailed in Section 3.
        \item Sample uniformly an element $r\in\Z_p^*$.
        \item If $m=1$, set $d = g^{\prod_{j\in[n]} a_j}$; otherwise set $d=g^r_n$. 
        \item Output ciphertext $c=(d,c_1,\ldots,c_\ell)$.
    \end{itemize}
    \item $\Dec(x,w)$, where $w$ is a witness that has some proposed
    solution $T\subset [\ell]$. Compute $c^* = \prod_{i\in T}c_i$.
    Output 1 if $c^* = d$ and output 0 otherwise.
    \item \textbf{Correctness:} Clearly if $T$ is indeed an exact
    cover, we know that 
    \[ \prod_{i\in T} c_i = \prod_{i\in T} g^{\prod_{j\in S_i} a_j}_{|S_i|} = g_n^{\prod a_i} \]
    \item \textbf{Security} This follows directly from Definition~\ref{DMNEA} below. In particular, consider the following reduction: given that no
    witness $w$ exists, if one could distinguish between $d =
    g^{\prod_{j\in[n]} a_j}$ and the random $g_n^r$, then this
    automatically breaks the Definition~\ref{DMNEA} assumption.
    
\end{itemize}
We will use this scheme, with a slight modification to a graded
encoding scheme that will be described below. In order to prevent 
\emph{zeroizing attacks}, one of the primary pitfalls that graded 
encoding schemes has had, we will use an asymmetric implementation of CLT13 \cite{clt13}, and set these levels to be vectors.

\subsection{Security Assumptions to Witness Encryption}

The following assumption will give us security for our scheme.
\begin{definition}[Decision Multilinear No-\text{Exact Cover} Assumption, \cite{witnenc_garg13}] \label{DMNEA}
Let $x= (n, S_1,\ldots,S_\ell)$ be an instance of the
\textsc{Exact Cover} problem that has no solution. Let the 
parameters of the group be sampled from
$\mathcal{G}(1^{\lambda + n}, n)$ with prime
order $p$. Let $a_1,\ldots,a_n,r$ be uniformly
random in $\Z_p$. For $i\in [\ell]$, let 
$h_i = (g_{|T_i|})^{\prod_{j\in T_i} a_j}$. It is 
hard to distinguish between
\[ (\texttt{params}, h_1,\ldots,h_\ell,g_n^{a_1,\ldots,a_n}) \text{ and } (\texttt{params}, h_1,\ldots,h_\ell,g_n^r).\]

\end{definition}

\section{Existing Multilinear Map/Graded Encoding Scheme Constructions}
While the existence of multilinear maps satisfying the above is still an open question, many candidates have been proposed, and subsequently broken. Multilinear maps are a special case of what are known as graded-encoding schemes.

\begin{definition}[Symmetric $\kappa$-Graded Encoding Scheme, \cite{GGH13, clt13}]
A $\kappa$-Graded Encoding System for a ring $R$ is a system of sets $\mathcal{S} = \{S_v^{(\alpha)} \subset \{0, 1\}^*\ :\ v \in \N, \alpha \in R\}$, where each set $S_v^{(\alpha)}$ represents the collection of $v$-level encodings of $\alpha$, with the following properties:
\begin{enumerate}
    \item For every $v \in \N$, the sets $\{S^{(\alpha)}_v: \alpha \in R\}$ are disjoint.
    \item There are binary operations $+$ and $-$ (on $\two^*$) such that for every $\alpha_1, \alpha_2 \in R$, every $v \in \N$, and every $u_1 \in S_v^{(\alpha_1)}$ and $u_1 \in S_v^{(\alpha_2)}$, it holds that $u_1 + u_2 \in S_v^{(\alpha_1 + \alpha_2)}$ and $u_1 - u_2 \in S_v^{(\alpha_1 - \alpha_2)}$ where $\alpha_1 + \alpha_2$ and $\alpha_1 - \alpha_2$ are addition and subtraction in $R$. 
    There is an associative binary operation $\times$ (on $\two^*$) such that for every $\alpha_1, \alpha_2 \in R$, every $v_1, v_2$ with $0 \leq v_1 + v_2 \leq \kappa$, and every $u_1 \in S_{v_1}^{(\alpha_1)}$ and $u_2 \in S_{v_2}^{(\alpha_2)}$, it holds that $u_1 \times u_2 \in S^{(\alpha_1 \cdot \alpha_2)}_{v_1 + v_2}$ where $\alpha_1 \cdot \alpha_2$ is multiplication in $R$.
\end{enumerate}
Graded encoding schemes, as outlined in GGH13 \cite{GGH13}, have a set of procedures they provide:

\begin{itemize}
    \item \textbf{Instance Generation.} $\InstGen(1^\lambda, 1^\kappa)$ takes as inputs $\lambda$, the security parameter, and $\kappa$, the required multilinearity level, and outputs $(\params, \pzt)$.
    \item \textbf{Ring Sampler.} $\samp(\params)$ outputs a level-0 encoding $a \in S^{(a)}_0$, where $\alpha$ is (nearly) uniform on $R$, though the encoding may not be uniform in $S_0^{(\alpha)}$
    \item \textbf{Encoding.} $\enc(\params, a)$ is a (possibly randomized, which in the case of CLT, it is) level-one encoding $u \in S^{(\alpha)}_1$, where $ a\in S^{(\alpha)}_0$.
    \item\textbf{Re-Randomization.} $\reRand(\params, i, u)$ re-randomizes encodings relative to the same level $i$, meaning given $u \in S^{(\alpha)}_i$, it outputs $u' \in S^{(\alpha)}_v$ such that, for any two $u_1, u_2 \in S^{(\alpha)}_v$, the output distributions of $\reRand(\params, i, u_1)$ and $\reRand(\params, i, u_2)$ are nearly the same.
    \item\textbf{Addition and negation.} For two encodings $u_1 \in S^{(\alpha_1)}_i$ and $u_2 \in S^{(\alpha_2)}_i$ at the same level $i$, we have $\add(\params, u_1, u_2) = u_1 + u_2 \in S^{(\alpha_1 + \alpha_2)}_i$ and $\mathsf{neg}(\params, u_1) = -u_1 = \in S^{(-\alpha_1)}_i$.
    \item\textbf{Multiplication.} For two encodings $u_1 \in S^{(\alpha_1)}_i$ and $u_2 \in S^{(\alpha_2)}_j$, $\mul(\params, u_1, u_2) = u_1 \times u_2 \in S_{i + j}^{(\alpha_1 \alpha_2)}$.
    \item \textbf{Zero-test.} The procedure $isZero$, with outputs in $\two$, satisfies $\isZero(\params, \pzt, u) = 1 \iff u \in S_{\kappa}^{(0)}$, i.e. it reveals whether a given encoding is a top-level encoding of 0.
    \item \textbf{Extraction.} $\mathsf{ext}(\params, \pzt, u)$ computes a random function of a ring element from their top-level encoding. 
\end{itemize}
\end{definition}
Multilinear maps can be viewed as a special case of graded encoding schemes, where the levels are given a group structure and where the encoding is deterministic. 

 For a multilinear map defined on the groups $G_1 \times G_2 \times \dots \times G_n$ with respect to generators $(g_1, \dots, g_n)$, simply take $R = \Z/|G|\Z$ and let $S^{(\alpha)}_v$ just be $g_v^{\alpha}$. The operation $+$ is then multiplication within a given group $G_v$, and $\times$ is then application of the multilinear map, which takes $g_{v_1}^{\alpha_1}, g_{v_2}^{\alpha_2} \mapsto g_{v_1 + v_2}^{\alpha_1\alpha_2}$.


GGH13 was the first proposal of a graded-encoding scheme \cite{GGH13}. We do not discuss the construction in detail, and instead focus on the construction used in our implementation, CLT13.

In the subsequent section, we outline portions of the originally proposed CLT13 construction relevant to our use and the break in CLHRS15 \cite{CHLRS15}, as well as some other breaks.

\subsection{CLT13 Graded Encoding Scheme Construction}
\subsubsection{Protocol}

\textbf{Overview:} We generate $n$ secret random primes $p_i$, compute $x_0 = \prod p_i$, and randomly sample some $z$ that is invertible modulo $x_0$. We will encode a message $\boldsymbol{m} \in \Z^n$ at level $k$ as 
\begin{align*}
    c \equiv \frac{r_i \cdot g_i + m_i}{z^k} \pmod{p_i}
\end{align*}
where $c$ is well-defined modulo $x_0$ via Chinese Remainder Theorem, $\{g_i\}$ are random small primes generated at the start of the protocol and the $r_i$ are small values generated for each encoding. Note that by construction, $m_i$ is well defined modulo $g_i$, so in fact the scheme encodes vectors from the ring $R = \Z_{g_1} \times \cdots \times Z_{g_n}$.

\textbf{Instance Generation.} For security parameter $\lambda$ and multilinearity level $\kappa$, we generate the following:
    \begin{itemize}
        \item $n$ secret random $\eta$-bit primes $p_i$. Let $x_0 = \prod_{i = 1} ^n p_i$. These $p_i$ can be considered \textit{ciphertext} moduli.
        \item A random integer $z$ that is invertible modulo $x_0$.
        \item $n$ random $\alpha$-bit primes $g_i$ (these are the \textit{plaintext} moduli).
        \item A secret matrix $\boldsymbol{A} = (a_{ij}) \in \Z^{n \times \ell}$, where each $a_{ij} \rfrom [0, g_i) \cap \Z$.
        \item An integer $y$, and three sets of integers $\{x_j\}_{j = 1} ^\tau$, $\{x'_j\}_{j = 1} ^\ell$, $\{\Pi_j\}_{j = 1} ^n$, a zero-testing vector $\pzt$, and a seed $s$ for a strong randomness extractor. Of relevance for us are only the integer $y$, the sets $\{x'_j\}_{j = 1} ^\ell$ and $\{x_j\}_{j = 1} ^\ell$, and the zero-testing vector $\pzt$, the definitions of which we will describe below.
    \end{itemize}  
We publish the parameters $\params = (n, \eta, \alpha, \rho, \beta, \tau, \ell, y, \{x_j\}_{j = 1} ^\tau, \{x'_j\}_{j = 1} ^\ell, \{\Pi_j\}_{j = 1} ^n, s)$.

\textbf{Sampling level-zero encodings}: $\{x'_j\}_{j = 0} ^\ell$ is a set of level-0 encodings of the column vector $\boldsymbol{a}_j \in \Z^n$ of the secret matrix $\boldsymbol{A}$. More precisely, we set
    \begin{align*}
        x'_j \equiv r'_{ij} \cdot g_i + a_{ij} \pmod{p_i},\qquad r'_{ij} \rfrom (-2^\rho, 2^\rho) \cap \Z
    \end{align*}
    where $x'_j$ is defined $x_0$ using Chinese Remainder Theorem. To generate a random level-0 encoding, we simply take a random subset sum of these $x'_j$, i.e. we generate a random binary vector $\boldsymbol{b} \in \two^\ell$ and output, for $\samp(\params)$, 
    \begin{align*}
        c = \sum_{i:\boldsymbol{b}_i = 1}x'_i \mod x_0
    \end{align*}
    which is then some encoding of $\boldsymbol{m} = \boldsymbol{A} \cdot \boldsymbol{b}$. The authors show that the distribution of this $\boldsymbol{m}$ is then close to uniform over $R$.
    
\textbf{Encoding at higher levels:} $y$ is a level-1 random encoding of $1$. Explicitly,
    \begin{align*}
        y \equiv \frac{r_i \cdot g_i + 1}{z} \pmod{p_i} \qquad r_i \rfrom (-2^\rho, 2^\rho) \cap \Z
    \end{align*}
    Now given a level-0 encoding $c$ of some $\boldsymbol{m} \in \Z^n$ (i.e. $c$ satisfying $c \equiv r_i' \cdot g_i + m_i \mod p_i$), $c_1 = c \cdot y \mod x_0$ is then a level-1 encoding:
    \begin{align*}
        c_1 \equiv \frac{(r'_i + m_i \cdot r_i + r_i \cdot r'_i \cdot g_i) \cdot g_i + m_i}{z} \pmod{p_i}
    \end{align*}
    Thus, we have that $\enc(\params, c)$ computes $c' = c \cdot y \mod x_0$.  More generally, to compute a level $k$ encoding, we compute $c \cdot y^k \mod x_0$.
    
    However, in cases where it is necessary to prevent recovery of $c$ given $c'$, we must also rerandomize $c'$ at its encoding level using $\reRand$. This is done by publishing public encodings of $0$ at level $k$, (for example, the $\{x_j\}$ are encodings of $0$ at level $1$), and adding them to $c'$, as in the case of generating random level-0 encodings, but with slight adjustments. We omit the specific details of the construction.
    
\textbf{Adding and Multiplying Encodings.} It is clear from the form of the encryptions that homomorphic addition and multiplication are possible, provided the numerators remain small enough.

\textbf{Zero-testing.} $\isZero(\params, \pzt, u_\kappa)$ outputs a value in $\two$, with $\isZero(\params, \pzt, u_\kappa) = 1 \iff u_\kappa \in S_\kappa^{(0)}$. This is done by constructing an appropriate integer matrix $\boldsymbol{H} = (h_{ij}) \in \Z^{n \times n}$. We then publish the \textit{zero-testing vector} $\pzt \in \Z^n$ satisfying
\begin{align*}
    (\pzt)_j = \sum_{i = 1} ^n h_{ij} \cdot (z^\kappa \cdot g_i^{-1} \mod p_i) \cdot \frac{x_0}{p_i}
\end{align*}
This enables us to test whether a top level encoding $c$ is zero or not, by computing $\omega = c \cdot \pzt \mod x_0$ and testing whether $\|\omega\|_\infty = \max_j |\omega_j| $ is less than $x_0 \cdot 2^{-\nu}$, where $\nu$ is another parameter used for the randomness extractor $\mathsf{ext}$, which we will not discuss. Intuitively, this works due to the fact that we can write $c$ as $c = q_i \cdot p_i + (r_i \cdot g_i + m_i) \cdot (z^{-\kappa}\mod p_i)$, for some $q_i \in \Z$, and then we have that
\begin{align}\label{succ-zero-test}
    (\boldsymbol\omega)_j &= (c \cdot \pzt \mod x_0)_j = \sum_{i = 1} ^n h_{ij} \cdot (r_i + m_i \cdot (g_i^{-1} \mod p_i)) \cdot \frac{x_0}{p_i} \mod x_0 \\
    &= \sum_{i = 1} ^n h_{ij} \cdot r_i \cdot \frac{x_0}{p_i} \mod x_0 \nonumber
\end{align}
Thus, when each $r_i$ is small and each $m_i$ is $0$, we have that $(\boldsymbol\omega)_j$ will be small, provided $\boldsymbol{H}$ is chosen appropriately, the details of which we omit.

\subsection{CLT13 Zeroizing attack}

Attacks against the original CLT13 encryption scheme as well as other graded encoding schemes surfaced about a year later and were predominantly based on exploiting the multiple public encodings of zero that are published by the protocol. This class of attacks, termed as \say{zeroizing} attacks, were demonstrated to be able to successfully recover all secret parameters used in the encryption scheme, rendering it unusable. 

\subsubsection{CLHRS15 \cite{CHLRS15}\ CHL${}^{+}$14 Zeroizing Attack}

Keeping to the notation used by the authors, we define the following: 
Let $\CRT_{(p_1,\cdots, p_n)}(r_1,\cdots, r_n)$ be the unique integer $x$ in the interval $(-\frac{1}{2}\prod_{i=1}^n p_i, \frac{1}{2}\prod_{i=1}^n p_i]$ that satisfies the congruence $x\equiv r_i\mod p_i$ for all $i\in[n]$. 
For succinctness we write $\CRT_{(p_i)}(r_i)$ to represent the above.

The attack is predicated upon finding solutions to a modified version of the approximate greatest common divisor problem:
\newcommand{\sampD}{\mathcal{D}_{\chi_\varepsilon, \eta, n}(p_1,\cdots, p_n)}
\begin{definition}[Sampleable $\mathsf{CRT-ACD}$ distribution \cite{CHLRS15}]
Let $n,\eta, \varepsilon\in \N$ and let $\chi_\varepsilon$ be some distribution over the integers in $(-2^\varepsilon,2^\varepsilon)$. Let $p_1,\cdots, p_n$ be $\eta$-bit primes and define $x_0=\prod_{i=1}^np_i$ as in CLT. Then the \emph{sampleable $\mathsf{CRT}$-$\mathsf{ACD}$ distribution} $\mathcal{D}_{\chi_\varepsilon,\eta, n}(p_1,\dots, p_n)$ is defined as 
\begin{align*}
    \sampD = \{\CRT_{(p_i)}(r_i)\mid \forall i,\; r_i\rfrom \chi_\varepsilon\}
\end{align*}
\end{definition}

\begin{definition}[$\CRT$-$\mathsf{ACD}$ with auxiliary input \cite{CHLRS15}]
Let $n,\eta, \varepsilon,\chi_\varepsilon, p_i,x_0$ be defined as above. Additionally define $\widehat{p}_i = \frac{x_0}{p_i}$. Then the $\CRT$-$\mathsf{ACD}$ with auxiliary input problem is the following:

\emph{Given the values $x_0$, $\widehat{P}=\CRT_{(p_i)}(\widehat{p}_i)$, and polynomially many samples from $\sampD$, determine the values of all $p_i$.}
\end{definition}

The solution-finding algorithm relies on the following lemma, proven in full in the paper:

\begin{lemma}\label{paper-lemma}
For a given auxiliary value $\widehat{P}=\CRT_{(p_i)}(\widehat{p}_i)$, any value $a=\CRT_{(p_i)}(r_i)$ drawn from $\sampD$ satisfies:
\begin{align*}
    a\cdot \widehat{P}\mod x_0 = \CRT_{(p_i)}(r_i\cdot \widehat{p}_i) = \sum_{i=1}^n r_i\cdot \widehat{p}_i
\end{align*}
as long as $\varepsilon +\log n + 1 <\eta$.
\end{lemma}

Using this lemma, the paper demonstrates how to retrieve the secret parameters $p_1,\cdots, p_n$ of the CLT encryption given access to polynomially many samples from $\sampD$. Given two draws $a=\CRT_{(p_i)}(a_i)$ and $b=\CRT_{(p_i)}(b_i)$, we have that $ab\widehat{P}\mod x_0=\sum_{i=1}^n a_ib_i\widehat{p}_i \mod x_0$. Supposing that the $a_i$'s and $b_i$'s are small enough such that the above lemma is satisfied. Then the right-hand side is equal to $\sum_{i=1}^n a_ib_i\widehat{p}_i$. This can be represented in matrix form as follows:
\begin{align*}
    ab\widehat{P} \mod x_0 = \begin{pmatrix}
    a_1 & \cdots &a_n
    \end{pmatrix}
    \begin{pmatrix}
    \widehat{p}_1 & \cdots & 0 \\
    \vdots & \ddots & \vdots \\
    0 & \cdots & \widehat{p}_n
    \end{pmatrix}
    \begin{pmatrix}
    b_1\\\vdots\\b_n
    \end{pmatrix}
\end{align*}
For the attack, we assume a draw of $2n+1$ samples from $\sampD$ with the following labels:
\begin{align*}
    a_i
    &= \CRT_{(p_k)}(a_{k.i}),\quad i\in [n] \\
    b
    &= \CRT_{(p_k)}(b_k) \\
    c_j
    &= \CRT_{(p_k)}(c_{k.j}),\quad j\in [n]
\end{align*}
It is worth emphasizing here that we do not have access to the remainders $a_{k,i},b_k,c_{k,j}$ themselves.
To ensure that the conditions of Lemma~\ref{paper-lemma} are satisfied, the parameters of the problem must additionally satisfy $3\varepsilon+\log n+1 <\eta$. 
Under those conditions, we define and compute the parameters $w_{i,j}, w'_{i,j}$ as the following, where equality is given by Lemma~\ref{paper-lemma}:
\begin{align*}
    &w_{i,j}
    = a_ibc_j \widehat{P}\mod x_0 = \sum_{k=1}^n a_{k,i}b_k c_{k,j} \widehat{p}_k 
    \qquad 1\leq i,j \leq n \\
    &w'_{i,j}
    = a_ic_j \widehat{P}\mod x_0 = \sum_{k=1}^n a_{k,i}c_{k,j} \widehat{p}_k 
    \qquad\quad\; 1\leq i,j \leq n 
\end{align*}
Using the matrix representation as presented earlier, the relations can be rewritten as
\begin{align*}
    \mathbf{W}
    &= \mathbf{A}^T\cdot\mathrm{diag}(b_1\widehat{p}_1,\cdots, b_n\widehat{p}_n)\cdot\mathbf{C} \\
    \mathbf{W'}
    &= \mathbf{A}^T\cdot\mathrm{diag}(\widehat{p}_1,\cdots, \widehat{p}_n)\cdot\mathbf{C}
\end{align*}
where the matrices are defined as $\mathbf{A}^T=(a_{k,i})$, $\mathbf{C}=(c_{k,j})$, $\mathbf{W}=(w_{i,j})$, and $\mathbf{W'}=(w'_{i,j})$, all elements of $M_{n\times n}(\Z)$.

Under the assumption that $\mathbf{A}$ and $\mathbf{C}$ are invertible over $\Q$, the following matrix is well-defined:
\begin{align*}
    \mathbf{V} 
    := \mathbf{W}\cdot(\mathbf{W'})^{-1} 
    = \mathbf{A}^T\cdot \mathrm{diag}(b_1,\cdots, b_n)\cdot (\mathbf{A}^T)^{-1}
\end{align*}
The eigenvalues of this matrix are then the set of remainders $\{b_1,\cdots, b_n\}$, which can be computed in polynomial time.
It is shown in the paper that as long as the $b_i$'s are distinct, it suffices to compute a few $\mathsf{GCD}$s to extract the values of all the secret primes $p_1,\cdots, p_n$. If either $\mathbf{A}$ or $\mathbf{C}$ is singular, the procedure can be repeated until that is not the case.

To apply this CLT13, we first review some of the public parameters:
\begin{itemize}
    \item $x_j = \CRT_{(p_i)}\left(\frac{r_{ij}g_i}{z}\right)$ for $j \in [\tau]$ are our public level-1 encodings of $0$ used for rerandomization.
    \item $x'_j = \CRT_{(p_i)}(x'_{ij})$, where $x'_{ij} = r'_{ij}g_i + a_{ij}$, with $r'_{ij} \rfrom (-2^\rho, 2^\rho) \cap \Z$ for $i \in [n]$, $j \in [\ell]$ are encodings of $\boldsymbol{a}_j$ at level-0.
    \item $y = \CRT_{(p_i)}\left(\frac{r_ig_i + 1}{z}\right)$, where $r_i \rfrom (-2^\rho, 2^\rho)\cap\Z$ for $i \in [n]$ is our public level-1 encoding of $1$.
    \item $(\pzt)_j = \sum_{i = 1} ^n h_{ij} \cdot (z^\kappa \cdot g_i^{-1} \mod p_i) \cdot \frac{x_0}{p_i} \mod x_0$ for $j \in [n]$.
\end{itemize}
Note that in $(\ref{succ-zero-test})$, we have that when $c$ is a top-level encoding of zero, and we write $c = \CRT_{(p_i)}(r_ig_i/z^\kappa)$, we will have that
\begin{align*}
    (\pzt)_ja = \sum_{i = 1} ^n \widehat{p}_ih_ir_i = \CRT_{(p_i)}(\hat{p}_i h_i r_i)
\end{align*}
where second equality is as in Lemma~\ref{paper-lemma}, provided it is smaller than $x_0/2$. We can construct such $c$ by either taking $x'_j \cdot x_1' \cdot x_k \cdot y^{\kappa - 1}$ or $x'_j \cdot x_k \cdot y^{\kappa - 1}$, for $1 \leq j,k \leq n$.

We omit the specific calculations, but this ultimately produces a set of equations similar to the case discussed for solving $\CRT-\mathsf{ACD}$ with auxiliary input, and enables us to solve for the specific $x'_{ij}$, which then give the $p_i$ by computing $\mathsf{GCD}(x'_1 - x_{i1}, x_0)$, fully breaking the scheme.

\subsection{Modifications to CLT13/other attacks}
Following the CHLRS15 attack, the CLT13 scheme was updated to the CLT15 \cite{CLT15} scheme, where the zero-testing procedure was modified to be non-linear in the moduli, thus preventing the quadratic form representation in the previous section.
The scheme was also modified so that the prime $x_0$ did not need to be published.
However, this modification was demonstrated in \cite{CLT15break} to exhibit similar vulnerabilities. It turns out that using the public parameters, the value of $x_0$ could be recovered, under which the original attack could then be mounted.

The attack in CHLRS15 \cite{CHLRS15} was extended in \cite{cryptoeprint:2014:930} to not specifically require the level-0 encodings of $0$ that are originally given publically in the parameters. As described in \cite{cryptoeprint:2015:596}, most rely on the ability to construct top-level encodings of $0$ and successful zero tests. 

As we will detail, our implementation of witness encryption will require no public encodings of zero (at level-$\kappa$ or level-0). However, there still exist attacks for this, such as in \cite{cryptoeprint:2014:929} that do not explicitly require encodings of zeros beyond at the top level either.

\subsection{Specifics of Our Implementation}\label{specifics}
The C implementation of CLT13 that we build upon has been modified to support asymmetric graded-encoding schemes, the definition of which we give below. Essentially, instead of having integer indices for the levels, we have vector indices; elements within the same level can still be added. Multiplying two elements produces an element that is indexed by the vector that is the sum of the original indices.
\renewcommand{\v}{\boldsymbol{v}}
\begin{definition}[$T$-Graded Encoding Scheme, \cite{GGH13}] Let $T \subset \N^\tau$ be a finite set (for some integer $\tau > 0$), and let $R$ be a ring. A $T$-Graded Encoding System for $R$ is a system of sets $\mathcal{S} = \{S^{(\alpha)}_{\boldsymbol{v}} \subset \two^* \mid v \in \Lambda(T), \alpha \in R\}$, with the following properties:
\begin{itemize}
    \item For every fixed index $\v \in \Lambda(T)$, the sets $\{S_{\v}^{(\alpha)} \mid \alpha \in R\}$ are disjoint (hence they form a partition of $S_{\v} := \bigcup_\alpha S_{\v}^{(\alpha)}$.
    \item There are binary operations `$+$' and `$-$' (on $\two^*$) such that for every $\alpha_1, \alpha_2$ in the same level $\v \in \Lambda(T)$, and every $u_1 \in S_{\v}^{(\alpha_1)}$ and $u_2 \in S_{\v}^{(\alpha_2)}$, it holds that
    \begin{align*}
        u_1 + u_2 \in S_{\v}^{(\alpha_1 + \alpha_2)}\text{ and }u_1 - u_2 \in S_{\v}^{(\alpha_1 - \alpha_2)}
    \end{align*}
    where $\alpha_1 + \alpha_2$ and $\alpha_1 - \alpha_2$ are addition and subtraction in $R$.
    \item There is an associative binary operation `$\times$' (on $\two^*$) such that for every $\alpha_1, \alpha_2 \in R$, every $\v_1, \v_2$, with $v_1 + v_2 \in \Lambda(T)$, and every $u_1 \in S_{\v_1}^{(\alpha_1)}$ and $u_2 \in S_{\v_2}^{(\alpha_2)}$, it holds that
    \begin{align*}
        u_1 \times u_2 \in S^{(\alpha_1 \cdot \alpha_2)}_{v_1 + v_2}.
    \end{align*}
    where $\alpha_1 \cdot \alpha_2$ is multiplication in $R$, and $\v_1 + \v_2$ is vector addition in $\N^\tau$.
\end{itemize}
$\Lambda(T)$ denotes the interior of $T$, i.e. the set of points $x$ in $T$ for which there exists $w \in T$ such that $w_i \geq x_i$ for all $i$, and there is one index $j$ where $w_j > x_j$.
\end{definition}

Furthermore, the CLT13 construction can be modified to support \textit{multiple} $z_i$'s instead of a single $z$ in order to make the underlying group structure asymmetric. Based on our understanding of the codebase (\cite{codeclt13}), an encoding of a message $\boldsymbol{m} \in \Z^{n}$ at level $\boldsymbol{v}$ is then
\begin{align*}
    c \equiv \frac{r_i \cdot g_i + m_i}{\prod_{j = 1} ^n z^{v_j}_j} \pmod{p_i}
\end{align*}

Following a call with Professor Amit Sahai (which we may or may not have fully understood), we implement our encryption, primarily based on Garg's original construction. For a given \textsc{Exact Cover} instance $x$ comprised of sets $S_1 \dots, S_\ell \in \mathcal{P}([n])$, we encrypt as follows:

$\mathsf{Enc}(x, m):$ 
\begin{itemize} 
    \item Sample $a_0, \dots, a_n$ uniformly and independently from $\Z_{g_1} \times \Z_{g_2} \times \cdots \Z_{g_n}$.
    \item Let $c_i$ be an encryption of $\prod_{j \in S_i} a_j$, encoded at level $\boldsymbol{v}_j \in \two^n$, where $(\boldsymbol{b}_j)_k = 1$ if $k\in S_i$ and $0$ otherwise. The codebase provides support for encoding single integers rather than $n$-dimensional integer vectors; it does so, from our understanding, by setting the value to the first component, and letting the rest be zero.
    \item Sample uniformly an element $ \in \Z^*_p$.
    \item Let $d_1$ be a level $\boldsymbol{v} = (1, 1, \dots, 1)$ encryption of $\prod_{j = 1} ^n a_i$. Let $d_0$ be a random element encoded at that level.
    \item Output $(d_m, c_1, \dots, c_n)$. Letting $x_0, \pzt, \nu$ be as in the definition of the CLT13 graded encoding scheme, we also output these three values.
\end{itemize}
$\mathsf{Dec}(x, w)$ takes $w$ is a witness that has some proposed solution $T \subset [\ell]$ of indices of subsets, and $x$, a ciphertext.
\begin{itemize}
    \item Compute $c^* = \prod_{i \in T} c_i$. Output 1 if $d - c^*$ passes the zero-test, and $0$ otherwise.
\end{itemize}
\textbf{Correctness.} If $T$ is indeed an \textsc{Exact Cover}, we have that $c^* = \prod_{i \in T} c_i$ will be an encoding at level $\sum_{i \in T}\boldsymbol{v}_i = (1, 1, \dots, 1)$ of the value $\prod_{i \in T} \prod_{j \in S_i} a_j = \prod_{i= 1} ^n a_i$, which is exactly a top-level encoding of $\prod_{i = 1} ^n a_i$. We then then compute $d - c^*$ and zero-test it, i.e. compute whether
\begin{align*}
    |(d - c^*) \cdot (\pzt)_j \mod x_0 | \leq x_02^{-\nu},
\end{align*}
where we take $(d - c^*) \cdot (\pzt)_j \mod x_0 $ to be in the range $(-x_0/2, x_0/2]$,
is small for all $j$. This test is possible because we give $x_0, \pzt, \nu$ publically, and these are the only parameters that zero-testing relies upon.

\textbf{Security.} Security is defined based on the hardness of a variant of Decision Multilinear No-Exact-Cover Problem for Graded Encoding schemes, where we explicitly define $\params$ to be the parameters given out in the encryption and change some attributes to fit our construction:
\begin{definition}[Decision Graded Encoding No-Exact-Cover Problem \cite{witnenc_garg13}]
Let $x = \{T_i \mid T_i \subset [n], i \in [\ell]\}$ be an instance of \textsc{Exact Cover} that has no solution. Let $\params = (x_0, \pzt, \nu)$ be the attributes for some $n$-graded encoding system as above. Generate $a_1, \dots, a_n, r$ as random level-0 encodings, and let $c_i$, $d_0$, $d_1$ be as above. Then the following are indistinguishable:
\begin{align*}
    (\params, c_1, \dots, c_n, d_1)\text{ and }(\params, c_1, \dots, c_n, d_0)
\end{align*}
We do not know whether or not this holds for CLT13, given the public parameters we output.
\end{definition}


\section{Code and Documentation}

The workflow is as follows: given an input \texttt{.txt} file containing a description of a Sudoku puzzle or a Pentomino tiling problem in a specified format, the Python script translate that into an exact set cover problem, printing out an encoding of that into a separate text file. 
Then the C code, built upon the modified CLT13 implementation by Carmer and Malozemoff \cite{codeclt13}, performs the encryption of the message based on the exact set cover problem.

The sections that follow describe the implementation details of each phase.

\textbf{File Structure}
\begin{itemize}
        \item \texttt{clt.c}
        \item \texttt{clt\_elem.c}
        \item \texttt{clt\_tree.c}
        \item \texttt{estimates.c}
        \item \texttt{utils.c}
\end{itemize}
The main file containing majority of the theoretical encoding described above lies in \texttt{clt.c}.

\subsection{Witness Encryption}
We will describe the code surrounding the witness encryption
with \textsc{Exact Cover}, and how exactly it relates to the
theory described in Section 3.4.

We generate each of the $a_i\in \Z_{x_0}$ as follows. 
\begin{cpp}
    mpz_t *a;
    mpz_t target_product;
    a = malloc(kappa * sizeof(mpz_t));
    mpz_init_set_ui(target_product, 1);
    mpz_array_init(*a, kappa, pow(2, lambda + 1));
    for (size_t i = 0; i < kappa; i++) {
        mpz_urandomb_aes(a[i], rng, 2 * lambda);
        mpz_mul(target_product, a[i], target_product);
    }    
\end{cpp}
From here we can start to use Carmer and Malozemoff's code in order to encode our vectors at specific level sets \cite{codeclt13}.
\begin{cpp}
int
clt_encode(clt_elem_t *rop, const clt_state_t *s, size_t n, const mpz_t *xs,
           const int *ix)
\end{cpp}
Here, we will encode into group element \cppinline{rop} $\in \Z_{x_0}^*$ the value of a vector \cppinline{xs} $\in \Z^n$. \cppinline{ix} is the level vector.

The \cppinline{clt\_state\_t} data structure to contains the $\params$ as described in Section 3.1.
\begin{cpp}
struct clt_state_t {
    size_t n;                   /* number of slots */
    size_t nzs;                 /* number of z's in the index set */
    size_t rho;                 /* bitsize of randomness */
    size_t nu;                  /* number of most-significant-bits to extract */
    mpz_t *gs;                  /* plaintext moduli */
    aes_randstate_t *rngs;      /* random number generators (one per slot) */
    mpz_t x0;
    mpz_t pzt;                  /* zero testing parameter */
    mpz_t *zinvs;               /* z inverses */
    union {
        crt_tree *crt;
        mpz_t *crt_coeffs;
    };
    size_t flags;
};
\end{cpp}
There is notation to be clarified here, the state is independent of any given iteration of an encoding. The $n$ passed as an argument to \cppinline{clt\_encode} is dimension of the encoded vector, while here it is the number of primes generated, and thus the maximum dimension of a vector (the code simply fills in unassigned dimensions as zeros). However, in our uses, they will be the same value. The following code is used to calculate the value of a given vector and place the resulting value in \cppinline{rop}:
\begin{cpp}
if (s->flags & CLT_FLAG_OPT_CRT_TREE) {
        mpz_t *slots = mpz_vector_new(s->n);
#pragma omp parallel for
        for (size_t i = 0; i < s->n; i++) {
            mpz_random_(slots[i], s->rngs[i], s->rho);
            mpz_mul(slots[i], slots[i], s->gs[i]);
            if (i < n)
                mpz_add(slots[i], slots[i], xs[i]);
            /* mpz_add(slots[i], slots[i], xs[slot(i, n, s->n)]); */
        }
        crt_tree_do_crt(rop->elem, s->crt, slots);
        mpz_vector_free(slots, s->n);
    } else {
        mpz_set_ui(rop->elem, 0);
#pragma omp parallel for
        for (size_t i = 0; i < s->n; ++i) {
            mpz_t tmp;
            mpz_init(tmp);
            mpz_random_(tmp, s->rngs[i], s->rho);
            mpz_mul(tmp, tmp, s->gs[i]);
            if (i < n)
                mpz_add(tmp, tmp, xs[i]);
            /* mpz_add(tmp, tmp, xs[slot(i, n, s->n)]); */
            mpz_mul(tmp, tmp, s->crt_coeffs[i]);
#pragma omp critical
            {
                mpz_add(rop->elem, rop->elem, tmp);
            }
            mpz_clear(tmp);
        }
    }
\end{cpp}
From here, our encryption scheme can output/publish all the encodings $c_i$ along with $d_{m}$. The top-level zero-test parameter is published at 
\begin{cpp}
struct clt_pp_t {
    mpz_t x0;
    mpz_t pzt;                  /* zero testing parameter */
    size_t nu;                  /* number of most-significant-bits to extract */
};
\end{cpp}
Thus with this data, any witness with a solution to our given
\textsc{Exact Cover} can decrypt a message. All that remains 
is to find a clever person who can solve the problem and can 
use the \cppinline{pzt} parameter to find the value of our 
encrypted bits.

In our actual encoding, we take a single \textsc{Exact Cover} problem and use it to encode multiple bits. Explicitly, we generate, for an Exact Cover problem with sets $S_1, \dots, S_\ell$ with entries in $[n]$, the respective $c_1, \dots, c_\ell$. We then encode $x \in \two^{*}$ bit by bit, in each case computing $d_0$ if $x_i = 0$, and $d_1$ otherwise (where $d_0$, $d_1$ are described in section~\ref{specifics}). This still has some semblance of encryption due to the fact that the encodings are randomized. That they are indistinguishable from encodings of other elements is not guaranteed by the scheme. We would have preferred to be able to encrypt each bit using a unique puzzle, but the encodings, even at smaller encryption sizes, have large size, and thus this is very difficult.

Since we encrypt a puzzle that we know to have a solution, no level of security is guaranteed, since we only have soundness security. 

\subsection{Reductions to \textsc{Exact Cover}}

\subsubsection{class \texttt{SudokuReduction}}
We first give a theoretical outline of the reduction: 

In Sudoku, the goal is to fill up a $9\times 9$ grid, such that each row, column and $3\times 3$ sub-grid contains the numbers 1-9 exactly once. The reduction from Sudoku to \textsc{Exact Cover} rests on the observation that the constraints of the problem can be succinctly stated as follows:
\begin{itemize}
    \item Each cell contains exactly one integer in the set $\{1,\cdots,9\}$.
    \item Each integer appears exactly once in every row.
    \item Each integer appears exactly once in every column.
    \item Each integer appears exactly once in every sub-grid.
\end{itemize}
Each constraint in the list above is instanced $81$ times in the puzzle. For example, the first applies to every pair $(r,c)$ corresponding to the coordinates of a cell. Likewise, the second applies to every pair $(r,v)$ of row index and value. A completed Sudoku grid is then a valid solution if those $4\times81=324$ constrains are all satisfied.

We begin by presenting how a blank Sudoku puzzle can be encoded, and the follow up with how to account for pre-filled cells.

For a blank Sudoku puzzle, we let the $324$ elements of the set-cover problem represent the constraints. For concreteness, let these elements be $\{p_1,\cdots, p_{81}, q_1,\cdots, q_{81}, r_1,\cdots r_{81}, s_1,\cdots, s_{81}\}$.
We let each set be associated with a triple $(r,c,v)$, which represents the assignment of the cell $(r,c)$ to the value $v$. Each such assignment implies 4 statements.
\begin{itemize}
    \item Cell $(r,c)$ contains one integer.
    \item Row $r$ contains one copy of $v$.
    \item Column $c$ contains one copy of $v$.
    \item The sub-grid containing $(r, c)$ contains one copy of $v$.
\end{itemize}
Therefore we can construct a map $F$ that sends each triple $(r,c,v)$ to some 4-element set $\{w,x,y,z\}$ that represent each of the statements above. Then the \textsc{Exact Cover} problem over these elements and sets enforces that the collection of all such statements satisfy the constraints of the Sudoku problem. 

For Sudoku puzzles with pre-filled cells, we transform the \textsc{Exact Cover} problem of a blank Sudoku to that of the particular puzzle by eliminating sets that correspond to conflicting assignments to the assignments already on the board (Say the top left corner is a '1'. Then we get rid of the sets $(1,1,2),\cdots, (1,1,9)$ since those correspond to assigning some other number to the top left cell, and so on).

We note that Sudoku can be extended to an $n^2\times n^2$ grid with analogous rule-sets, and this is supported by our implementation of the reduction. 

\begin{itemize}
    \item \begin{python}
    def __init__(self, puzzle):
        self.puzzle = puzzle
        self.n = len(puzzle)
        self.b = int(self.n ** 0.5)
        self.collection = [[0] for _ in range(self.n ** 3)]
        self.gen_sets()
        self.trim_sets()
    \end{python}
    
    The puzzle is a $n\times n$ grid of numbers that are either the digits $0$ to $n-1$ or the keyword \texttt{None} which indicates that it is initially empty. $n$ is the side-length of the puzzle and $b$ is the side-length of any one of the subgrids. $\texttt{gen\_sets()}$ and $\texttt{trim\_sets()}$ are functions that implement the Sudoku-to-exact-cover reduction, and their implementations are below:
    
    \item \begin{python}
    def gen_sets(self):
        for row, col, val in itertools.product(range(self.n), range(self.n), range(self.n)):
            offset = self.n ** 2
            st = [0 for _ in range(4 * self.n ** 2)]

            st[self.n * row + col] = 1
            st[offset + row * self.n + val] = 1
            st[offset * 2 + col * self.n + val] = 1
            st[int(offset * 3 + ((row // self.n ** 0.5) * self.n **
                   0.5 + (col // self.n ** 0.5)) * self.n + val)] = 1

            self.collection[row * self.n ** 2 + col * self.n + val] = st
    \end{python}
    
    Based on the reduction, given a triple of values $(r,c,v)$ representing the cell at row $r$, column $c$ having value $v$, this corresponds to the row numbered $rn^2+cn+v$ in the constraint table. This row is initialized with $4n^2$ 0's, partitioned into 4 blocks of $n^2$ entries. A $1$ is inserted in exactly one entry of each block to represent that selecting this row (which in turn represents inserting value $v$ in the cell $(r,c)$) satisfies this constraint. This is as follows
    \begin{enumerate}
        \item The first block represents the constraint that each cell is only allowed to contain one value.
        \item The second block represents the constraint that a value can only appear once in each row.
        \item The second block represents the constraint that a value can only appear once in each column.
        \item The second block represents the constraint that a value can only appear once in each minor grid.
    \end{enumerate}
    This updates the \texttt{self.collection} value.
    
    \item \begin{python}
    def trim_sets(self):
        bad_indices = set()
        for row in range(self.n):
            for col in range(self.n):
                if self.puzzle[row][col] is None:
                    continue
                bad_indices.update(set(map(lambda v: self.to_index(row, col, v), range(self.n))))
                bad_indices.update(set(map(lambda c: self.to_index(row, c, self.puzzle[row][col]), range(self.n))))
                bad_indices.update(set(map(lambda r: self.to_index(r, col, self.puzzle[row][col]), range(self.n))))
                bad_indices.update(set(map(lambda x: self.to_index((row // self.b) * self.b + x/ 
                 self.b, (col // self.b) * self.b + x // self.b, self.puzzle[row][col]), range(self.n))))
                bad_indices.remove(row * self.n ** 2 + col * self.n + self.puzzle[row][col])

        for i in bad_indices:
            self.collection[i] = [0 for _ in range(4 * self.n ** 2)]
    \end{python}
    
    This function takes as input the initial state of the Sudoku puzzle and deletes all the sets that are incompatible with the sets that are enforced by that initial state.
    
    \item \begin{python}
    def to_index(self, row, col, val):
        return int(row * self.n ** 2 + col * self.n + val)
    \end{python}
    
    Handles the conversion from a triple $(r,c,v)$ to the corresponding index in the constraint table.
    
    \item \begin{python}
    def to_exact_cover_sets(self):
        def bin_to_numeric(ls):
            return [i for i, b in enumerate(ls) if b]
        subsets = list(map(bin_to_numeric, self.collection))
        return subsets, 4 * self.n ** 2
    \end{python}
    
    This function converts each row of the constraint matrix to a set of four numbers representing their elements, and returns the list of conversions along with the total number of unique elements to cover (324 in the case of standard Sudoku, as noted in above.
\end{itemize}

\subsection{class \texttt{PentominoBitmasks}}

This reduction is slightly less involved since the constraints are easier to represent. 

Suppose there are $5n$ cells arranged in some manner on a grid and $n$ Pentomino tiles. Then the goal of Pentomino tiling is to find an arrangement of the $n$ tiles such that each cell on the board is covered by exactly one tile. 

The elements in the corresponding exact set cover can be taken to be the set $\{c_1,\cdots, c_{5n}, p_1,\cdots, p_n\}$, where the $c$'s correspond to some labelling of the cells and the $p$'s correspond to some labelling of the Pentominos. Each set then has 6 elements and takes the form $\{c_{i_1},c_{i_2}, c_{i_3}, c_{i_4}, c_{i_5}, p_i\}$, where $p_i$ is a certain Pentomino and the $c_i$'s are a set of cells that can potentially be covered by that Pentomino. It is clear that the \textsc{Exact Cover} problem encodes the constraints in the original tiling problem.

In the classic Pentomino tiling problem, the board has exactly 60 cells and there is exactly one of each type of Pentomino. However, this can also be generalized to arbitrary board sizes and sets of available Pentominos (with multiplicity).

\begin{itemize}
\item 
    \begin{python}
    def __init__(self, puzzle, pieces):
        self.puzzle = puzzle
        self.n = len(puzzle)
        self.m = len(puzzle[0])
        self.pieces = pieces
        self.piece_count = sum(pieces)
        self.cart_to_lbl = None
        self.tile_bitmasks = PentominoBitmasks().bitmask_list
        self.cells = 0
        self.init_cart_to_lbl()
        self.collection = []
        self.gen_sets()
    \end{python}
    The key parameters are as follow:
    \texttt{puzzle} is a $n \times m$ of integers in $\{1, 0\}$, where a '1' indicates that the cell is empty.
    \texttt{pieces} is a length 12 array containing the number of pieces of each Pentomino $\{O, P, Q, R, S, T, U, V, W, X, Y ,Z\}$.
    We use the Conway labeling scheme for Pentominos.
    
    \texttt{cart\_to\_lbl} is a dictionary that maps cartesian coordinates to integer labels.
    \texttt{collection} eventually holds the exact set cover description.
    \texttt{init\_cart\_to\_lbl()} generates the cell labeling and \texttt{gen\_sets()} performs the exact set cover translation. Their implementations are detailed below.

\item
    \begin{python}
    def init_cart_to_lbl(self):
        counter = 0
        cart_to_lbl = dict()
        for r, row in enumerate(self.puzzle):
            for c, cell in enumerate(row):
                if cell == 1:
                    cart_to_lbl[(r,c)] = counter
                    counter += 1
        self.cart_to_lbl = cart_to_lbl
        self.cells = len(self.cart_to_lbl)
    \end{python}
    This function keeps an incrementing counter and labels the empty cells in row-major order.
    
\item
    \begin{python}
    def gen_sets_for_piece(self, pc_id):
        sets = []
        for mask in self.tile_bitmasks[pc_id]:
            maskn = len(mask)
            maskm = len(mask[0])
            for r in range(self.n-maskn+1):
                for c in range(self.m-maskm+1):
                    overlap = False
                    thisset = []
                    for i in range (maskn):
                        for j in range (maskm):
                            if mask[i][j] == 0:
                                continue
                            elif mask[i][j] == 1 and self.puzzle[i+r][j+c] == 0:
                                overlap = True
                                break
                            else:
                                thisset.append(self.cart_to_lbl[(i+r, j+c)])
                        if overlap:
                            break
                    if not overlap:
                        sets.append(thisset)
        return sets
    \end{python}
    This function takes as input a label of a Pentomino piece, and generates a list of 5-element sets. Each set consists of the labels of 5 cells that can be covered by one piece corresponding to \texttt{pc\_id}.

\item
    \begin{python}
    def gen_sets(self):
        ctr = 0
        for id in range(len(self.pieces)):
            templ = self.gen_sets_for_piece(id)
            for _ in range(self.pieces[id]):
                chunk = list(map(lambda x : x+[self.cells + ctr], templ))
                self.collection += chunk 
                ctr += 1
    \end{python}
    This function loops over the set of Pentominos that are part of the puzzle (accounting for multiplicity). For the $i$-th puzzle piece, we append the element \texttt{cells}$+i$ to each of the 5-element sets returned by \texttt{gen\_sets\_for\_piece\_id}, corresponding to the translation described in above.
\end{itemize}

\section{Encryption using a Sudoku Puzzle}
We demonstrate the usablility of our implementation by encoding three Bitcoin wallets with the three Sudoku problems listed below.

\begin{figure}[t]
    \centering
    \includegraphics[width = 3cm]{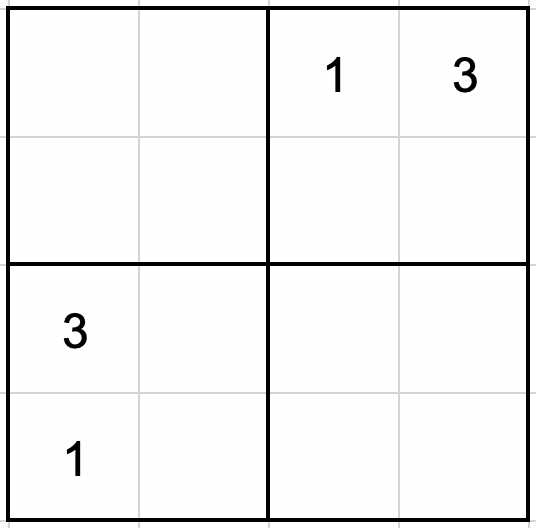}
    \caption{An \textit{unsolvable} sudoku puzzle. The way to uncover the message is to break our implementation. }
    \label{unsolvable}
\end{figure}

\begin{figure}[t]
    \centering
    \includegraphics[width = 5cm]{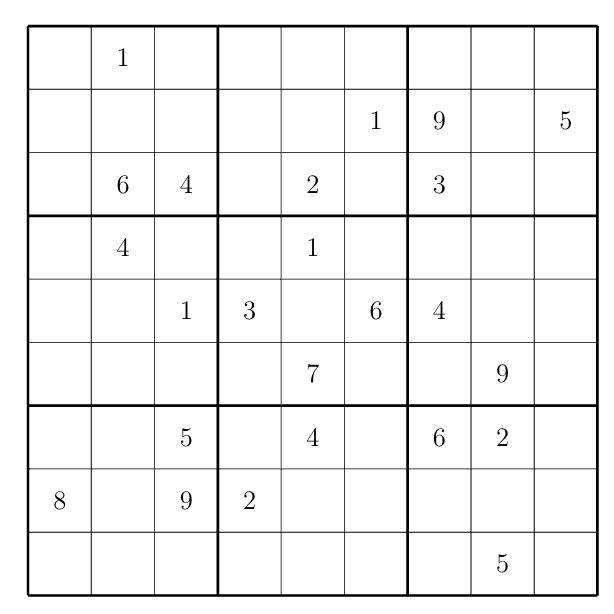}
    \caption{A solvable sudoku puzzle}
    \label{sudoku-1}
\end{figure}

\begin{figure}[H]
    \centering
    \includegraphics[width = 5cm]{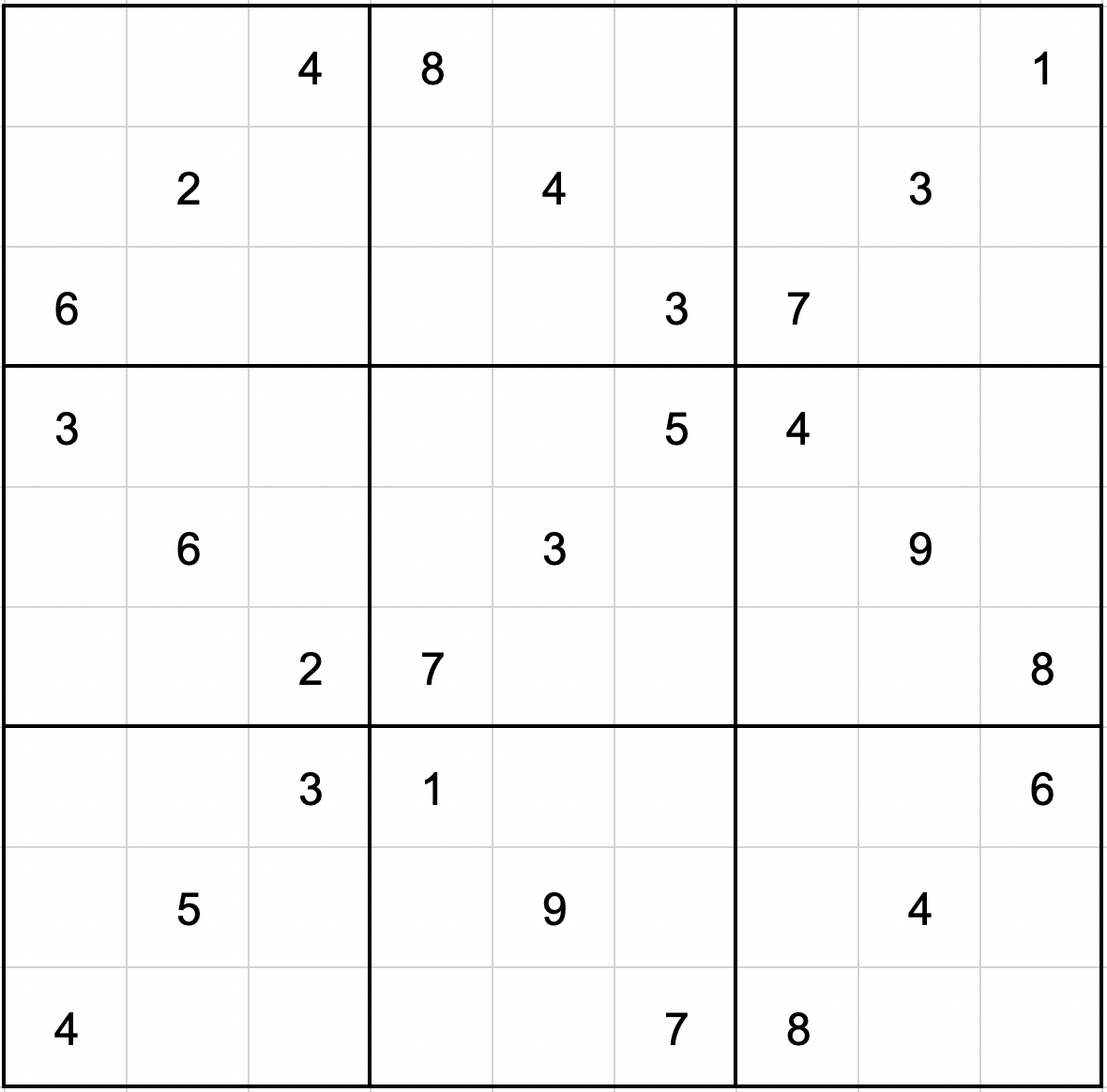}
    \caption{Another solvable sudoku puzzle}
    \label{sudoku-2}
\end{figure}

The encoding of the puzzles can be found at the link \href{https://github.com/guberti/witness-encryption-demos}{here}, along with instructions on how to load the puzzles and decrypt them. 

Sudoku 2 was encrypted with security parameter $\lambda = 20$, an intentionally low security parameter (and implicitly $\kappa = 248$, which is determined by the number of clues). Sudokus 1 and 3 are encrypted with $\lambda = 60$ - a more reasonable security parameter (though still lower than what is regularly used in private key encryption).

For the first clever individual
to solve the puzzle, we have encrypted the 256 bits of randomness used to generate a private key for a wallet containing a large number of Satoshi. In order to decrypt the bits, one should clone a copy of our code (also hosted \href{https://github.com/guberti/witness-encryption-demos}{here}, more detailed instructions in the README) onto their local repository in order to instantiate the graded encoding scheme and reduce their solution into an element of the ring $c^* \in \Z_{x_0}$. 

The Sudoku in figures 2 and 3 are connected to Bitcoin wallets with 2,770 Satoshi (in reference to Harvard's CS 227). This is worth about \$1.10 at writing. The unsolvable Sudoku in figure 1 is our grand prize, as the funds can only be retrieved through breaking our scheme. This wallet contains 27,700 Satoshi, worth \$11 at writing. 

\begin{center}
    \emph{Solve the puzzles and the prize could be yours!}
\end{center}

\newpage
\bibliography{bibliography.bib}
\bibliographystyle{plain}

\end{document}